\documentclass[runningheads]{llncs}
\usepackage[T1]{fontenc}
\usepackage{algorithm}
\usepackage{algpseudocode}
\usepackage{booktabs}

\usepackage{hyperref}
\usepackage{graphicx}
\usepackage{multirow}

\usepackage{amsmath,amsfonts,bm}

\def\eqref#1{equation~\ref{#1}}

\def\1{\bm{1}}

\def\vc{{\bm{c}}}

\def\ve{{\bm{e}}}

\def\vx{{\bm{x}}}

\def\mC{{\bm{C}}}

\def\mE{{\bm{E}}}

\def\mT{{\bm{T}}}

\DeclareMathAlphabet{\mathsfit}{\encodingdefault}{\sfdefault}{m}{sl}
\SetMathAlphabet{\mathsfit}{bold}{\encodingdefault}{\sfdefault}{bx}{n}

\def\gD{{\mathcal{D}}}

\def\gG{{\mathcal{G}}}

\def\gI{{\mathcal{I}}}

\def\gL{{\mathcal{L}}}

\def\gP{{\mathcal{P}}}

\def\gR{{\mathcal{R}}}

\def\gT{{\mathcal{T}}}
\def\gU{{\mathcal{U}}}

\def\gW{{\mathcal{W}}}

\begin{document}
%
\title{Graph Neural Patching for Cold-Start Recommendations}
%
%
\author{Hao Chen\inst{1} \and Yu Yang\inst{2} \and Yuanchen Bei\inst{3} \and
Zefan Wang\inst{4} \and
\\Yue Xu\inst{5} \and Feiran Huang\inst{4}}
\authorrunning{Hao Chen et al.}
%
\institute{
The Hong Kong Polytechnic University, Hong Kong SAR, China 
\and
The Education University of Hong Kong, Hong Kong SAR, China
\and
Zhejiang University, Hangzhou, China
\and
Jinan University, Guangzhou, China
\and
Alibaba Group, Hangzhou, China
\\
\email{sundaychenhao@gmail.com}, \email{yangyy@eduhk.hk}}
\maketitle              
\begin{abstract}
The cold start problem in recommender systems remains a critical challenge. Current solutions often train hybrid models on auxiliary data for both cold and warm users/items, potentially degrading the experience for the latter. This drawback limits their viability in practical scenarios where the satisfaction of existing warm users/items is paramount.
Although graph neural networks (GNNs) excel at warm recommendations by effective collaborative signal modeling, they haven't been effectively leveraged for the cold-start issue within a user-item graph, which is largely due to the lack of initial connections for cold user/item entities. Addressing this requires a GNN adept at cold-start recommendations without sacrificing performance for existing ones.
To this end, we introduce \underline{G}raph \underline{N}eural \underline{P}atching for Cold-Start Recommendations (GNP), a customized GNN framework with dual functionalities: GWarmer for modeling collaborative signal on existing warm users/items and Patching Networks for simulating and enhancing GWarmer's performance on cold-start recommendations. 
Extensive experiments on three benchmark datasets confirm GNP's superiority in recommending both warm and cold users/items.

\end{abstract}
\section{Introduction}
Recommender systems aim to retrieve deserving items for users out of massive online content and provide personalized recommendations~\cite{chen2024macro,guo2017deepfm,he2020lightgcn}. 
With countless new user registrations and item introductions occurring every second, these systems are frequently confronted with cold-start challenges, which is how to make recommendations for new users and items without historical interactions~\cite{chen2022generative,huang2023aligning,wei2021contrastive}. 
In a typical cold-start scenario for user-item recommendations, the system can access the auxiliary information for both warm and cold users/items to alleviate the cold-start problem~\cite{huang2024large,lee2019melu,wu2022survey}.
    
A primary challenge in cold-start recommendations is to effectively cater to new users and items while ensuring that the recommendation quality for existing users and items remains unaffected, given the significant presence of both in the system.
Given pre-trained warm embeddings and the auxiliary information, representative state-of-the-art models like DropoutNet \cite{volkovs2017dropoutnet} and Heater \cite{zhu2020heater} cultivate hybrid models to recommend for both cold users/items and warm users/items. 
These hybrid models are tasked with handling two distinctly different types of inputs concurrently: 
(i) For warm recommendations, they utilize \textit{warm embeddings} and the auxiliary information to craft warm representations.
(ii) For cold-start recommendations, they rely on \textit{placeholder embeddings} and the auxiliary information to produce a representation for new entities. 
While these hybrid cold-start models exhibit some proficiency in handling new entities, they often come at the cost of diminished performance for existing warm users and items. This trade-off occurs because the models must contend with handling two distinctly different types of inputs: the generic placeholder embeddings for newcomers and the pre-trained warm embeddings for the established base. 
The hybrid models struggle to deliver satisfactory recommendations for established users and items, which in turn limits the applicability of these solutions in real-world recommender systems. 

Central to the effectiveness of a warm recommender system is the capability to accurately model the interactions among existing users and items. In this regard, Graph Neural Networks (GNNs), renowned for their adeptness in handling graph-structured data, have secured a leading position in numerous recommendation scenarios with effective collaborative filtering~\cite{he2020lightgcn,wang2019ngcf,zhao2019intentgc}.
However, the direct application of current GNN models to address the cold-start problem is impeded by three primary challenges:
\begin{itemize}
\item Cold-start users and items lack historical interactions, which implies an absence of neighboring nodes within the graph structure. GNN models are unable to perform neighborhood aggregations, which are essential for refining the embedding representations of cold-start users and items, as they lack these neighboring connections.
\item Obtaining precise representations for these new entities is inherently challenging. Secondly, crafting a GNN that can sustain recommendation performance for existing users and items while also effectively handling cold-start recommendations is far from a trivial task.
\item Because of their recursive message passing mechanism, GNN models typically require more computational time compared to straightforward inner-product functions. Consequently, deploying GNN models for cold-start recommendations might face significant computational challenges, potentially outweighing the efficiency of current inner-product-based cold-start models.
\end{itemize}

To surmount the aforementioned challenges, this paper introduces a novel Graph Neural Patching model (GNP) designed to deliver satisfactory recommendations for both existing and cold-start users and items. The GNP framework consists of two distinct yet interconnected models: GWarmer and the Patching Network.
GWarmer is a specially crafted, efficient GNN model that enhances recommendations for existing users and items. Notably, GWarmer maintains a simplicity akin to inner-product calculations during inference, rendering the GNP framework more efficient than state-of-the-art cold-start models. Additionally, the Patching Network, which acts as a complementary module to GWarmer, provides cold-start recommendations by generating embeddings compatible with GWarmer from auxiliary information. In contrast to conventional cold-start models, the Patching Network is dedicated solely to cold-start recommendations, thereby ensuring that they do not disrupt the performance of the warm recommendation models.
Extensive experiments on three benchmark datasets, with three representative types of embedding methods, present that GNP statistically significantly outperforms other recommendation models on both hybrid recommendation performance~(including both warm and cold recommendation targets) and pure-warm recommendation performance with better efficiency. In summary, our contributions are listed as below:
\begin{itemize}
    \item This paper pioneers the integration of cold-start recommendations within GNN models, culminating in enhanced recommendation performance across the board for both warm and newly encountered cold users and items.
    \item This paper addresses the trade-offs inherent in hybrid cold-start models and introduces Graph Neural Patching (GNP), a solution that effectively delivers cold-start recommendations without detracting from the performance of recommendations for existing warm users and items.
    \item Extensive experiments on three benchmark datasets, with three different embedding models, show that GNP significantly outperforms other cold-start models on all situations with statistically significant margins.
\end{itemize}

\section{Related Works}
\subsection{Graph Neural Networks for Recommendations}
Graph Neural Networks (GNNs) have become increasingly prominent in many relations modeling tasks~\cite{bei2023reinforcement,hamilton2017graphsage,kipf2016gcn,wu2019sgc}. For the domain of recommender systems, GNNs provide an effective framework for collaborative filtering to model the intricate user-item relationships~\cite{chen2024macro,sharma2024survey,wu2022graph}.
GNNs for collaborative filtering operate by constructing a graph where nodes represent users and items, and edges represent interactions such as ratings or views~\cite{chen2024feedback,zhang2024multi}. The power of GNNs in this context stems from their capability to aggregate and transform node features through neural message passing, capturing both local and global dependencies within the graph. This allows the model to learn nuanced user and item embeddings that reflect their latent factors, which are critical for accurate recommendations. Representatively, NGCF~\cite{wang2019ngcf} and LightGCN~\cite{he2020lightgcn} construct bipartite graphs from historical user-item interactions and then embed the users and items with the help of graph convolutional networks~\cite{hamilton2017graphsage,kipf2016gcn}.

\subsection{Cold-Start Recommendations}
The cold-start problem is a quintessential challenge in recommender systems, where the system must make personalized suggestions for new (cold) users or items that lack historical interaction data~\cite{chen2022generative,huang2024large}. Traditional recommendation approaches, which rely heavily on past behavior, struggle with these new entities~\cite{rendle2012bprmf,rendle2020neural}. To tackle this issue, various strategies have been developed, such as utilizing auxiliary data like user demographics or item attributes, employing content-based filtering, and leveraging recommender models to predict preferences based on limited information. However, existing models mainly adopt separate modules for cold and warm recommendations respectively~\cite{volkovs2017dropoutnet,zhu2020heater}. In this way, their effectiveness is frequently offset by a decline in recommendation performance for the warm users and items. 

\begin{figure}[tbp]
\centering
\includegraphics[width=1.01\columnwidth , trim=1.6cm 0.4cm 1.2cm 1.7cm,clip]{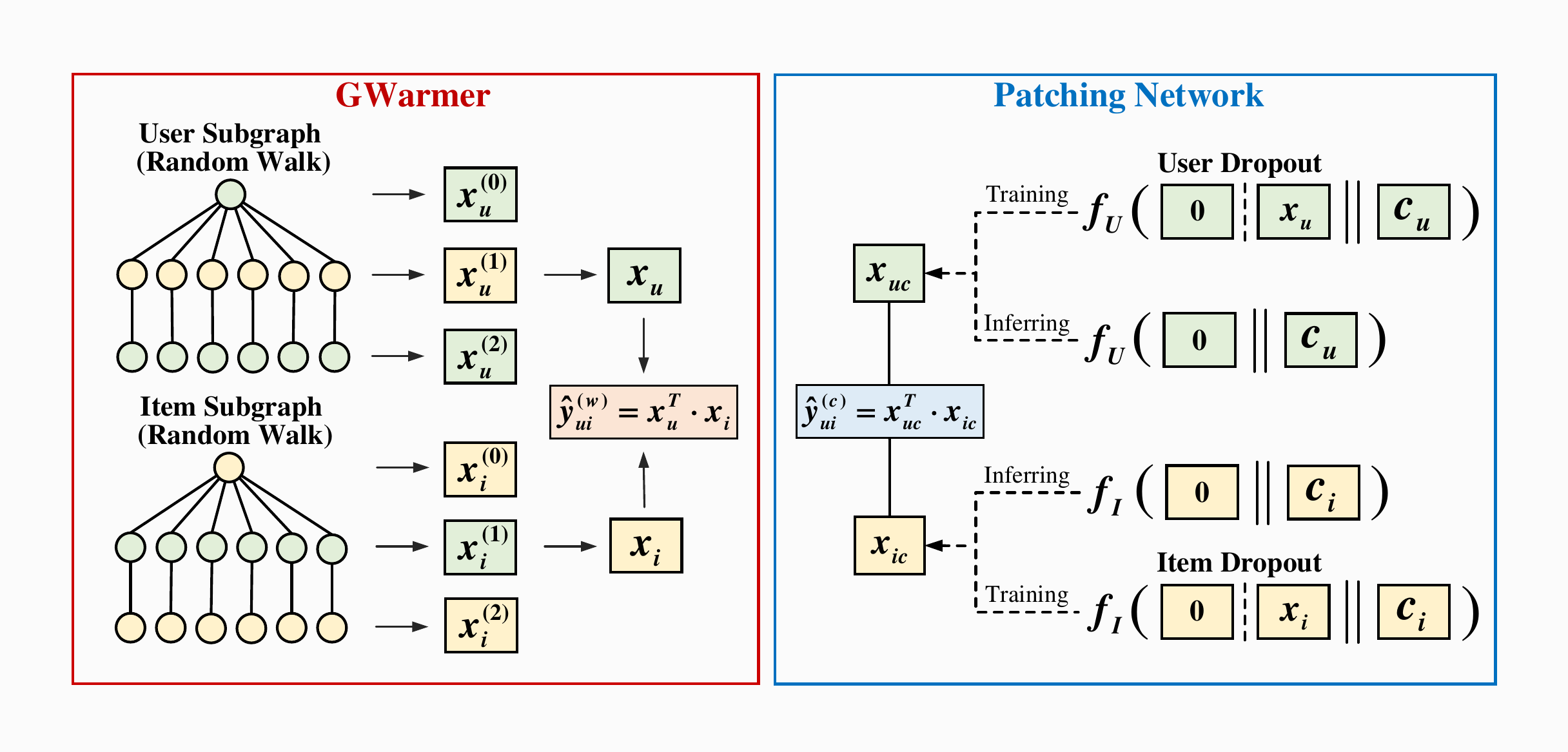}
\caption{The overall architecture of our proposed GNP. GWarmer provides recommendation for existing warm users and items, while the Patching Network flexibly recommend for cold-start users and items.}
\label{fig:framework}
\end{figure}

\section{Methodology}
This section first presents the whole framework of our Graph Neural Patching (GNP) model, and then we introduce the details of the proposed GWarmer and the Patching Network.
	
\subsection{Overall Framework}
Let $u\in\gU$ denote the user index, and $i\in\gI$ denote the item index. $\gG$ denotes the user-item bipartite graph constructed from the user-item interactions $\gR$. $\gU_w$ and $\gI_w$ denote the existing~(warm) user/item set. $\mC_U$ and $\mC_I$ denote the auxiliary information for all users/items. $\mE_U$ and $\mE_I$ denote the user/item embedding matrix trained by any given user-item embedding model.

We first define the general form of the warm/cold-start recommendation task. \emph{Warm recommendations}, by definition, mean that recommend warm items to warm users. For warm users and items, we can access the warm embeddings $\mE_U$ and $\mE_I$ and the historical interaction graph $\gG$. Formally, GWarmer predicts the relevance of a warm user-item pair $(u,i)$ with $\gW(u,i;\gG,\mE_U,\mE_I)$. By contrast, \emph{Cold-start Recommendations} refer to the recommendation for either cold-start user or cold-start item since the recommendation model cannot access the historical interaction graph and the embeddings for both users and items simultaneously. Without the graph and the embeddings, Patching Networks generate their cold-start embeddings from their auxiliary information $\vc_u$ and $\vc_i$, namely, $\gP(u,i;\vc_u,\vc_i)$. After defining GWarmer and Patching Networks, for an arbitrary user-item pair, GNP computes its relevance scores with the following formula,
\begin{align}
\hat{y}_{ui} = 
    \left\{
    \begin{aligned}
    &\gW(u,i;\gG,\mE_U,\mE_I) , \quad &u\in \gU_w \text{ and } i \in \gI_w , \\
    &\gP(u,i;\vc_u,\vc_i), &u \notin \gU_w \text{ or } i \notin \gI_w,
    \end{aligned}
    \right.
\end{align}
where $\gW(\cdot)$ denotes GWarmer, and $\gP(\cdot)$ denotes Patching Networks. We will detail these two parts in the next two subsections.

\subsection{GWarmer}
\label{sec:gwarmer}
The prevalent models~\cite{he2020lightgcn,huang2017label,rao2015grmf,rendle2012bprmf}, predominantly concentrate on developing more sophisticated user and item embeddings. These embeddings are subsequently utilized to forecast relevance scores through an inner product operation, as delineated below:
\begin{equation}
\hat{y}_{ui}^{(w)} = \ve_u^\top \ve_i.
\label{eq:inner}
\end{equation}

In alignment with this, an exemplary warm recommendation model tailored for practical recommender systems should possess the capability to refine any existing warm user-item embeddings by integrating them with the graph structure $\gG$. This enhancement should be achievable without the need to dictate the specific embedding models used, encompassing a range of methods such as Matrix Factorization (MF)~\cite{koren2009MF}, MetaPath2Vec (M2V)~\cite{dong2017metapath2vec}, and LightGCN~\cite{he2020lightgcn}. Furthermore, the model should be characterized by its efficiency in both the training phase and the inference process, ensuring that it meets the demands of real-time recommendation systems.

As depicted in Figure \ref{fig:framework}, the GWarmer model initiates its operation by constructing $K$-layer subgraphs for both warm users and items, leveraging the random-walk technique. This method effectively captures the intricate connections within the graph. Subsequently, the model aggregates these random-walk sets, thereby encapsulating the topological essence of the warm users and items within the graph structure. In the final stage, a self-adaptive inner product relevance function is designed to perform warm recommendations.
	
\textbf{Random-Walk Step.} 
The random-walk technique stands out as an efficient approach for capturing the essence of graph topology~\cite{grover2016node2vec,perozzi2014deepwalk}. As delineated in Figure \ref{fig:framework}, for any designated warm user or item $t$, we systematically initiate $S$ random walks, each spanning a length of $K$, with $t$ established as the origin node. Here, $S$ signifies the extent of our sampling endeavor. We collectively refer to this set of random walks as $\gT_K(t)$.
Note that in the context of bipartite user-item graphs, user nodes are interconnected with item nodes, and vice versa.
To provide a clear visual distinction in Figure \ref{fig:framework}, we employ green circles to represent user nodes and yellow circles to represent item nodes.

\textbf{Walk Pooling Step.} 
Upon acquiring the set of random walks, we proceed by replacing the node identifiers with their respective embedding vectors. Subsequently, we employ a mean pooling aggregator~\cite{gunets,hamilton2017graphsage}, to consolidate the random-walk set. Mean pooling is permutation-invariant on set elements, such that the order of the walks will not influence the aggregation. After the mean pooling aggregation, for node $t$ we get its layer-wise representation for each layer of the subgraph. Formally, the layer-wise representation is computed using the following equation:
\begin{equation}
    \vx_t^{(k)} =\frac{1}{S}\sum_{\mT_i\in\mathcal{T}_K(v)}\mT_{i}^{(k)},
\end{equation}
where $k$ denotes the layer number of the subgraph and $S$ denotes the sampling size of the random walk. Specifically, $\vx_t^{(0)}=E_t$.
	
 \textbf{Self-adaptive Inner Product}. 
Through the aforementioned two steps, we have successfully generated $K$ distinct representation vectors for each warm user or item $t$. Taking user $u$ as an example, $\vx_u^{(0)}$  is her own embeddings, while $\vx_u^{(1)}$ contains the mean embedding of her interacted items, and $\vx_u^{(2)}$ aggregates the embeddings of the users that have shared interacted items with $u$. 
This multi-layered approach provides a comprehensive profile of user $u$ from various perspectives.
Inspired by this multidimensional representation, we introduce a self-adaptive weighted sum mechanism that assigns unique weights to each layer.
This further enhances the flexibility and extensibility of the inner product. By tailoring the user and item representations with distinct weights, the relevance function is articulated as follows:
\begin{gather}
    \vx_u = \sum_k w_u^{(k)}\vx_u^{(k)}, \  \vx_i = \sum_k w_i^{(k)}\vx_i^{(k)}, \label{eq:warm}\\
    \hat{y}_{ui}^{(w)} =  \gW(u,i;\gG,\mE_U,\mE_I) = \vx_u^\top\vx_i,
\end{gather}
where $\{w_u^{(0)},\cdots,w_u^{(K)}\}$ and $\{w_i^{(0)},\cdots,w_i^{(K)}\}$ denotes the self-adaptive weights for users' and items' layer-wise representations, respectively. $\vx_u$ and $\vx_i$ denotes the GWarmer representation of $u$ and $i$.

In contrast to other recursive message passing models, GWarmer emerges as an exceptionally suitable candidate for large-scale recommender systems. The reasons for its suitability are manifold: Firstly, the random walk step and the walk pooling step are devoid of trainable parameters, which allows for their offline computation and storage in a parallelized fashion.
Secondly, in large commercial systems where historical user-item interactions are typically housed in distributed databases, GWarmer's approach offers a distinct advantage. Unlike traditional recursive GNNs that demand considerable time for recursively expanding neighbors, GWarmer's random walk policy substantially reduces the I/O overhead. 
Finally, once the GWarmer representations are stored, the model's operation during the online inference process is as straightforward as computing inner product functions. This simplicity makes GWarmer highly compatible with existing online recommender frameworks, such as Faiss~\cite{johnson2019faiss}.

\subsection{Patching Network}
Serving as a novel patch of GWarmer, the Patching Network needs to be consistent with GWarmer when conducting cold-start recommendation tasks. 
To ensure seamless integration, as depicted in Figure \ref{fig:framework}, we facilitate the connection between GWarmer and the Patching Network by leveraging GWarmer's representations in lieu of raw embeddings for the training of the Patching Network.
As shown in~\autoref{fig:framework}, we employ the popular cold-start dropout mechanism~\cite{volkovs2017dropoutnet,zhu2020heater} to optimize the Patching Networks by randomly masking the GWarmer representations of warm users/items with ratio $\tau$. 
For any given warm node $t$, we generate a random variable $p$, which follows a Bernoulli distribution with parameter $\tau$. The resulting masked GWarmer representation is articulated by the equation:
\begin{equation}
    \gD(\vx_t, \tau) =  p\cdot \mathbf{0} +  (1-p)\cdot \vx_t,
\end{equation}
where $\mathbf{0}$ is a zero vector that has the same dimension as $\vx_t$. 
Subsequently, for a given warm user-item pair $(u,i)$, the Patching Network will predict its relevance scores using the following formula:
    \begin{gather}
    \vx_{uc} = f_U(\gD(\vx_u,\tau)\parallel \vc_u), \ \vx_{ic} = f_I(\gD(\vx_i,\tau)\parallel \vc_i), \label{eq:patcher}\\
    \hat{y}_{ui}^{(c)}=\gP(u,i;\vc_u,\vc_i) = \vx_{uc}^\top\vx_{ic},
\end{gather}
where $f_U$ and $f_I$ denote the mapping function for users and items, respectively. Note that, $p$ is set as 1 during the inference of cold-start recommendation tasks when we set the embeddings of cold-start users/items as the placeholder embedding \textbf{0}.

\textbf{Objective Function}. In the GNP framework, the self-adaptive weights in Eq.~(\ref{eq:warm}) and the mapping functions in Eq.~(\ref{eq:patcher}) require optimization. Thus, we employ the Mean Squared Error~(MSE) loss~\cite{volkovs2017dropoutnet,zhu2020heater} to optimize both GWarmer and the Patching Network simultaneously. Formally, the loss function contains two parts and we present it as:
\begin{equation}
\gL = \sum_{\gR^+\cup\gR^-}(y_{ui} - \gW(u,i;\gG,\mE_U,\mE_I))^2 + (y_{ui} - \gP(u,i;\vc_u,\vc_i))^2,
\end{equation}
where $\gR^+$ denotes the observed user-item interactions, while $\gR^-$ denotes the sampling negative interactions. Thus, $y_{ui}=1$ for $(u,i) \in \gR^+$ and $y_{ui}=0$ for $(u,i) \in \gR^-$.

\section{Experiments}
In this section, we aim to answer the following two research questions: \textbf{RQ1}: How does GNP perform compared with state-of-the-art cold-start recommendation models? \textbf{RQ2}: What is the efficiency of the GNP framework? \textbf{RQ3}: How do the hyper-parameters, such as the dropout ratio $\tau$, affect GNP's performance?

\subsection{Experimental Settings}

\paragraph{\textbf{Dataset Description.}}
We conduct our experiments on three cold-start recommendation datasets~(CiteULike, XING, and WeChat). CiteULike~\cite{zhu2020heater} is a user-article dataset, where each article has a 300-dimension TF-IDF vector. XING~\cite{zhu2020heater} is a user-view-job dataset where each job is described by a 2738-dimension vector. WeChat is a user-video dataset, where each video can be represented by its NLP descriptions with a 200-dimension vector. We summarize the statistics of these three datasets in~\autoref{tab:dataset}. 
Here we mainly study the cold-start problem of items. For CiteULike and XING, 20\% of items are selected to be the cold-start items. Excluding the cold-start items, 65\% of historical interactions are used for embedding training, while 15\% of historical interactions are used for training GNP, and the remaining 20\% interactions are used for validation/testing. For WeChat, we split the dataset according to its timeline, where we use the newly produced videos as the cold-start items. 
We adopt the popular all-ranking evaluation protocol~\cite{he2020lightgcn,wang2019ngcf,zhu2020heater}. Specifically, we report Precision@K, Recall@K, and NDCG@K metrics, where K=20 for CiteULike and XING and K=100 for WeChat.

\begin{table}[htbp]
\small
\centering
\caption{Statistics of the three experimental datasets.}
\setlength{\tabcolsep}{2.7mm}{
 \begin{tabular}{ccccc}
 \toprule
 Dataset & \# User  & \# Item & \# Interaction & Density \\
 \midrule
 CiteULike & 5,551  & 16,980 & 204,986 & 0.22\% \\
 XING  & 106,881 & 20,519 & 3,856,580 & 0.18\% \\
 WeChat & 87,772 & 446,408 & 17,768,997 & 0.05\% \\
 \bottomrule
 \end{tabular}%
	}
	\label{tab:dataset}
\end{table}

\paragraph{\textbf{Baselines \& Embedding Methods.}} We consider two SOTA cold-start recommendation models: DropoutNet~\cite{volkovs2017dropoutnet} and Heater~\cite{zhu2020heater} as our baselines.
To evaluate the universality of GNP, we choose three different types of embedding methods: 
\begin{itemize}
\item MF~\cite{rendle2012bprmf} optimizes the matrix factorization with a pairwise ranking loss for recommendations, which is tailored to learn from implicit feedback.
\item M2V~\cite{dong2017metapath2vec} adopts metapath-based random walks on user-item graphs as corpus and then leverages skip-gram~\cite{mikolov2013word2vec} models to compute node embeddings. 
\item LightGCN~\cite{he2020lightgcn} utilizes GCN layers to aggregate neighbors' information, which presents to be currently the most powerful embedding model for user-item recommendation.
\end{itemize}

\paragraph{\textbf{Implementation.}}
The embedding size is fixed to 200 for all models, and all the embedding methods are implemented with the official codes. We skip LightGCN embeddings on the WeChat dataset since LightGCN reports out-of-CUDA memory errors on an NVIDIA 2080Ti GPU~(11G). In terms of GNP, we restrict the number of subgraph layers to 3. For computational efficiency, we sample the neighbors to accelerate the training and inference process. We set the sampling size of trajectory size as 25. We use Adam optimizer with a learning rate of 0.001, where the batch size is 1024. The coefficient of $\ell_2$ normalization is set as $1e^{-5}$, and the randomized dropout ratio $\tau$ is set as 0.5. Tanh is used as the activation function. The size of the hidden layers and the output layer is set as 200, while the depth of MLP is set as 2 by default. Moreover, an early stopping strategy is performed by observing the AUC scores.

\subsection{Main Results~(RQ1)}
\label{sec:results}
\autoref{tab:unified} reports the performance comparison of GNP and baselines on the hybrid recommendation tasks, which evaluates the ability of cold-start models to rank top-k warm and cold-start items out of all items. We conduct our experiments on three different warm embeddings. From this table, we observe that GNP statistically outperforms all other baselines on all embeddings and all datasets with $p<0.05$, demonstrating the consistent superiority of our GNP on hybrid recommendations. This further verifies that separating warm recommendation and cold-start recommendation models can exploit more potentialities from warm/cold-start recommendation models at the same time, resulting in better hybrid recommendation performance.
    
Further, to verify the pure-warm recommendation performance, in~\autoref{tab:warm}, we present how powerful are GNP and other baselines in recommending the top-k warm items out of all warm items. From this table, we highlight that GNP outperforms DropoutNet and Heater with statistically significant margins on all embeddings and all datasets. This further presents the necessity of solving cold-start problems with powerful GNN models.

\begin{table}[tbp]
  \centering
  \caption{Comparison of hybrid recommendation performance. The percentages give the relative improvement over the best baseline (underlined). OOM denotes out-of-memory during the model training. The superscript * denotes statistical significance against the best baseline with p<0.05.}
  \resizebox{\linewidth}{!}{
    \begin{tabular}{c|c|ccc|ccc|ccc}
    \toprule
    \multirow{2}[2]{*}{Embedding} & \multirow{2}[2]{*}{Model} & \multicolumn{3}{c|}{CiteULike} & \multicolumn{3}{c|}{XING} & \multicolumn{3}{c}{WeChat} \\
          &       & Recall & Precision & NDCG  & Recall & Precision & NDCG  & Recall & Precision & NDCG \\
    \midrule
    \multirow{4}[4]{*}{MF} & DropoutNet & 0.0780 & 0.0359 & 0.0654 & 0.1988 & 0.0712 & 0.1686 & 0.0519 & 0.0108 & 0.0315 \\
          & Heater & \underline{0.0968} & \underline{0.0479} & \underline{0.0849} & \underline{0.2190} & \underline{0.0792} & \underline{0.1921} & \underline{0.0640} & \underline{0.0130} & \underline{0.0403} \\
          & GNP & \textbf{0.1404} & \textbf{0.0668} & \textbf{0.1187} & \textbf{0.2346} & \textbf{0.0866} & \textbf{0.2055} & \textbf{0.0659} & \textbf{0.0133} & \textbf{0.0443} \\
\cmidrule{2-11}          & \textit{Improv.} & 45.04\%* & 39.46\%* & 39.81\%* & 7.12\%* & 9.34\%* & 6.98\%* & 2.97\%* & 2.31\%* & 9.93\%* \\
    \midrule
    \multirow{4}[4]{*}{M2V} & DropoutNet & \underline{0.1019} & \underline{0.0443} & \underline{0.0798} & \underline{0.2159} & \underline{0.0782} & \underline{0.1764} & 0.0685 & 0.0126 & 0.0401 \\
          & Heater & 0.0779 & 0.0406 & 0.0687 & 0.1946 & 0.0716 & 0.1658 & \underline{0.0770} & \underline{0.0139} & \underline{0.0481} \\
          & GNP & \textbf{0.1292} & \textbf{0.0565} & \textbf{0.1172} & \textbf{0.2541} & \textbf{0.0937} & \textbf{0.2204} & \textbf{0.0804} & \textbf{0.0154} & \textbf{0.0535} \\
\cmidrule{2-11}          & \textit{Improv.} & 26.79\%* & 27.54\%* & 46.87\%* & 17.69\%* & 19.82\%* & 24.94\%* & 4.42\%* & 10.79\%* & 11.23\%* \\
    \midrule
    \multirow{4}[4]{*}{LightGCN} & DropoutNet & 0.0933 & 0.0429 & 0.0795 & 0.2072 & 0.0756 & 0.1796 & OOM   & OOM   & OOM \\
          & Heater & \underline{0.1136} & \underline{0.0534} & \underline{0.1018} & \underline{0.2411} & \underline{0.0873} & \underline{0.2057} & OOM   & OOM   & OOM \\
          & GNP & \textbf{0.1631} & \textbf{0.0763} & \textbf{0.1411} & \textbf{0.2621} & \textbf{0.0954} & \textbf{0.2296} & OOM   & OOM   & OOM \\
\cmidrule{2-11}          & \textit{Improv.} & 43.57\%* & 42.88\%* & 38.61\%* & 8.71\%* & 9.28\%* & 11.62\%* & --   & --   & -- \\
    \bottomrule
    \end{tabular}%
    }
  \label{tab:unified}
\end{table}%

\begin{table}[ht]
\small
\centering
\caption{Comparison of warm recommendation tasks with the NDCG@K metric.}
\resizebox{0.78\columnwidth}{!}{
	\setlength{\tabcolsep}{2mm}{
\begin{tabular}{ccccc}
\toprule
   Embeddings   & Models   & CiteULike & XING & WeChat \\
\midrule
MF   & DropoutNet & 0.0737 & 0.2167 & 0.0394 \\
      & Heater & \underline{0.0934} & \underline{0.2653} & \underline{0.0489} \\
      & GNP & \textbf{0.1428} & \textbf{0.2757} & \textbf{0.0533} \\
      \cmidrule{2-5} & \textit{Improv.} & 52.89\%* & 3.92\%* &  9.00\%* \\
\midrule
M2V  & DropoutNet & 0.0765 & 0.2211 & 0.0567 \\
      & Heater & \underline{0.0857} & \underline{0.2393} & \underline{0.0629} \\
      & GNP & \textbf{0.1652} & \textbf{0.3157} & \textbf{0.0696} \\
      \cmidrule{2-5} & \textit{Improv.} & 92.77\%* & 31.93\%* &  10.65\%* \\
\midrule
LightGCN & DropoutNet & 0.0796 & 0.2506 & OOM \\
      & Heater & \underline{0.1232} & \underline{0.2930} & OOM \\
      & GNP & \textbf{0.1800} & \textbf{0.3398} & OOM \\
      \cmidrule{2-5} & \textit{Improv.} & 46.10\%* & 15.97\%* &  -- \\
\bottomrule
\end{tabular}%

	}}
	\label{tab:warm}
\end{table}

\begin{table}[ht]
\small
\centering
\caption{Comparison of average inferring time.}
\resizebox{0.58\columnwidth}{!}{
	\setlength{\tabcolsep}{2mm}{

\begin{tabular}{cccc}
\toprule
Model & CiteULike & XING  & WeChat \\
\midrule
DropoutNet & 56 s   & 79 s    & 1259 s\\
Heater & 73 s   & 97 s   & 1483 s\\
\midrule
GNP (ours) & 43 s   & 64 s   & 1054 s\\
\bottomrule

\end{tabular}%

	}}
	\label{tab:inferring}
\end{table}

\subsection{Efficiency Study (RQ2)}
In terms of the evaluation of model efficiency, we present the average inference times for DropoutNet, Heater, and GNP across the three datasets in question. As illustrated in Table \ref{tab:inferring}, GNP demonstrates superior efficiency compared to both DropoutNet and Heater. This advantage can be attributed to the streamlined nature of GWarmer during the inference phase, where it operates with the simplicity of an inner product calculation. Moreover, the Patch Network is a lightweight architecture, which is efficient during representation patching. In contrast, DropoutNet and Heater necessitate the generation of embeddings through a more complex multi-layer perceptron (MLP) function, which inherently requires additional computational resources and time.

    \begin{figure}[tbp]
		\centering
		\includegraphics[width=\columnwidth, trim=0cm 0cm 0cm 0cm,clip]{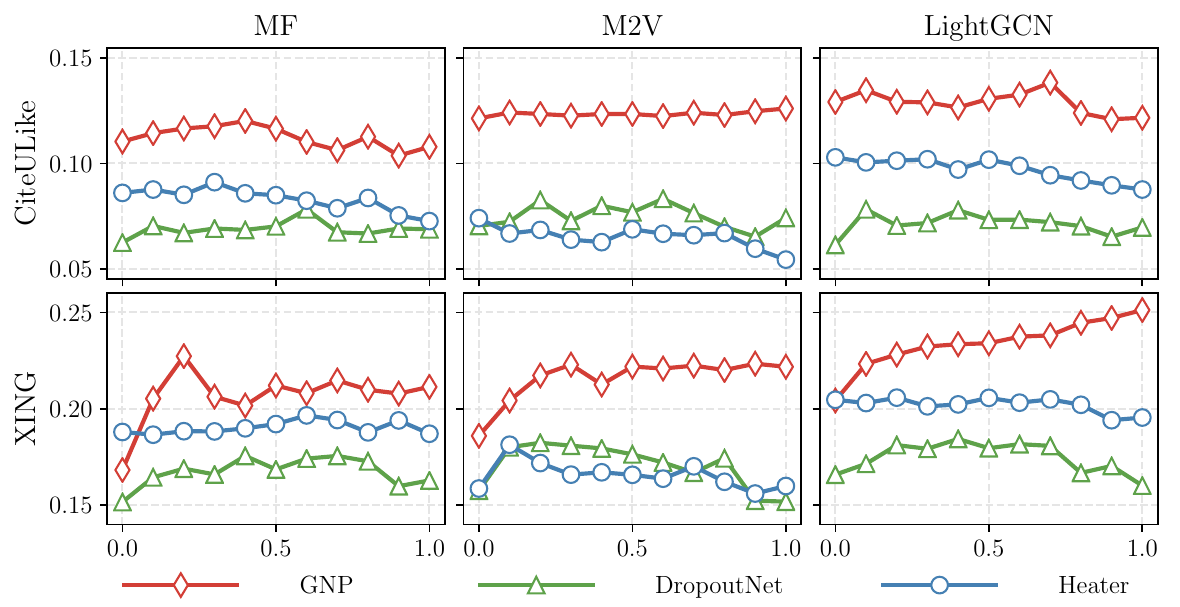}
		\caption{Comparison of different training dropout ratio $\tau$.}
         \label{fig:dropout}
	\end{figure}

\subsection{Ablation Study~(RQ3)}

In~\autoref{fig:dropout}, we present a comparative analysis of the hybrid recommendation performance between GNP, DropoutNet, and Heater, while varying the dropout ratio $\tau$ during training.
Observing from this figure, GNP consistently outperforms its counterparts, DropoutNet and Heater, across the majority of situations. This observation further reinforces the consistent superiority of GNP in handling hybrid recommendation tasks.
Especially, due to the Patching Network within GNP is specifically designed to address cold-start recommendations. Therefore, it is plausible that GNP achieves its optimal performance when the dropout ratio $\tau$ is set to 1. This configuration ensures that the model can fully leverage the Patching Network's capabilities for cold-start items while maintaining its effectiveness for warm items.

\section{Conclusion \& Future Works}
In this paper, we propose a simple but effective cold-start framework named Graph Neural Patching (GNP), which enhances the cold-start ability for GNN models with efficient modules (GWarmer and Patching Network). 
Specifically, GWarmer is an efficient GNN-based collaborative filtering module for the representation learning of warm users and items. Then, the crafted Patching Network is flexibly adopted to bolster the cold-start capabilities of GWarmer.
From the results of extensive experiments, GNP is generally powerful on all datasets and embeddings in both the hybrid recommendation and the warm recommendation. 

Moreover, the Patching Network that we propose serves as a versatile and efficient framework, enabling researchers to seamlessly integrate robust warm recommendation models with other effective cold-start models. This opens up a plethora of opportunities for future research endeavors, which could include: (i) designing more powerful patching cold-start models; (ii) flexibly combing with state-of-the-art warm recommendation models; (iii) investigating specialized patching models, such as large language models.

%
%
%
\bibliographystyle{splncs04}
\bibliography{reference}
\end{document}